\documentstyle[12pt]{article}

\begin{document}
\input{psfig.sty}
\begin{flushright}
\baselineskip=12pt
UPR-947-T \\
\end{flushright}

\begin{center}
\vglue 1.5cm
{\Large\bf GUT Breaking on $M^4 \times T^2/(Z_2 \times Z_2^{\prime})$}
\vglue 2.0cm
{\Large Tianjun Li~\footnote{E-mail: tli@bokchoy.hep.upenn.edu,
phone: (215) 898-7938, fax: (215) 898-2010.}}
\vglue 1cm
{ Department of Physics and Astronomy \\
University of Pennsylvania, Philadelphia, PA 19104-6396 \\  
U.  S.  A.}
\end{center}

\vglue 1.5cm
\begin{abstract}
We construct two $SU(5)$ models on the space-time
$M^4 \times T^2/(Z_2 \times Z_2^{\prime})$  where the gauge and Higgs
fields 
are in the bulk and the Standard Model fermions are on the 
brane at the fixed point or line.
For the zero modes, the SU(5) gauge symmetry is broken down
to  $SU(3) \times SU(2) \times U(1) $ due to non-trivil orbifold
projection.
In particular, if we put the Standard Model fermions on the 3-brane at the
fixed point
in Model II, we only have the zero modes and KK modes of the
Standard Model gauge fields and two Higgs doublets on the observable
3-brane.
 So, we can have the low energy unification, and solve 
the triplet-doublet splitting problem,
 the gauge hierarchy problem, and the proton decay problem.
\\[1ex]
\\[1ex]

\end{abstract}

\vspace{0.5cm}
\begin{flushleft}
\baselineskip=12pt
July 2001\\
\end{flushleft}
\newpage
\setcounter{page}{1}
\pagestyle{plain}
\baselineskip=14pt

\section{Introduction}
Grand unified theory (GUT) gives us a simple
 and elegant understanding of the quantum numbers of the quarks and
leptons,
and the success of gauge coupling unification in the MSSM strongly support
 this idea. Although the GUT at high energy scale has
become widely accepted now, there are some problems in GUT:  the grand
unified
gauge symmetry breaking mechanism, the triplet-doublet splitting problem
and
 the proton decay problem.

A new scenario to explain above questions in GUT has been suggested 
by Kawamura~\cite{kaw1, kaw2, kaw3},
 and further discussed by Altarelli, Barbieri, Feruglio, Hall,
Hebecker, Kawamoto,
Normura, and March-Russell~\cite{AF, HN, kk, HMR, bhn1N, bhn1, HMRN}.
 The key point is that the GUT
gauge symmetry exists in 5 or higher dimensions and is broken down to the
Standard Model gauge symmetry for the zero modes due to non-trivial 
orbifold projection on the multiplets and gauge generators in GUT. 
The attractive models have been constructed explicitly, where
the supersymmetric 5-dimensional SU(5) models are broken down to
the $N=1$ supersymmetric Standard Model through the compactification on
$S^1/(Z_2\times Z_2')$. The GUT symmetry breaking and triplet-doublet
splitting problems have been solved neatly by the orbifold projection.

In this letter, we will discuss the GUT with SU(5)~\cite{GG}
gauge symmetry breaking on the space-time 
$M^4 \times T^2/(Z_2 \times Z_2')$ where the $M^4$ is the 4-dimensional
Minkowski space-time. 
We assume that the SU(5) gauge fields and two 5-plet Higgs fields
are in the bulk and the Standard Model fermions are on the observable
brane which is located at the fixed point or line.
For simplicity, we do not consider
supersymmetry here.

We shall present two models in detail. The first
model is the generalization of previous models~[$1-10$]. Because $T^2$ is
homeomorphic to $S^1 \times S^1$, the extra space orbifold in this model
 is in fact $S^1/(Z_2 \times Z_2') \times S^1/(Z_2 \times Z_2')$
where two $Z_2$ and two $Z_2'$ act simultaneously, and
the observable brane is located at one of the 3 orbifold fixed points. 
 If the Standard Model fermions were on the
3-brane which is located at $(0, 0)$ or $(0, \pi R_2)$, the gauge group
will be $SU(5)$,
then, the SM fermions must form the full $10^i + \bar 5^i$ $SU(5)$
multiplets. In order
to avoid the proton decay, the GUT scale ($1/R_1$ or $1/R_2$) must be the
order of
 $10^{15\sim 16}$ GeV, and we have to put the two Higgs 5-plets in the
bulk to
give the GUT scale masses to the Higgs triplets. So, there may exist the
gauge hierarchy problem. And
if the Standard Model fermions were on the brane which is 
located at $(\pi R_1/2, \pi R_2/2)$,
the gauge group will be $SU(3)\times SU(2)\times U(1)$. If we considered
the
SM fermions form the full $10^i + \bar 5^i$ $SU(5)$ multiplets, the proton
may decay
by exchange $A_5^{\hat a}$ or $A_6^{\hat a}$, so, the GUT scale must be at
order of $10^{15\sim 16}$ GeV and we have to put the two Higgs 5-plets in
the bulk.
Thus, we might have the gauge hierarchy problem. If we considered the
SM fermions do not form the full $SU(5)$ multiplets and they only preserve
the
$SU(3)\times SU(2)\times U(1)$ gauge symmetry, the proton decay problem
will be
avoided because the SM fermions do not couple to the gauge fields
related to the broken gauge generators. 
In addition, we can put the two Higgs doublets on the observable brane
instead of putting
two Higgs 5-plets in the bulk. Therefore, the extra dimensions might be
large 
and the gauge hierarchy problem might be solved in the mean
time~\cite{AADD}.

In the second model, the extra space orbifold is indeed $S^1/Z_2
\times S^1/Z_2'$. If the Standard Model fermions were on the 4-brane which
is located at
$y=0$, or $z=0$, or $y=\pi R_1/2$, or $z=\pi R_2/2$,
the gauge symmetry will be $SU(3)\times SU(2)\times U(1)$. And the 
phenomenological discussions will be similar to those in Model I when we
put
the SM fermions on the 3-brane at $(\pi R_1/2,\pi R_2/2 )$, except that,
if
we considered the SM fermions form the full $10^i + \bar 5^i$ $SU(5)$
multiplets,
we will double the generations and each generation in the SM comes from
the zero modes of
two generations on the 4-brane due to the projection operator $P^y$ or
$P^z$.
If the Standard Model fermions were on the 3-brane which is located
at any fixed point, the gauge symmetry will be
$SU(3)\times SU(2)\times U(1)$. In addition, on the observable 3-brane,
 we project out not only the zero 
modes of the gauge fields related to the broken $SU(5)$ gauge generators
and the
triplet Higgs fields, but also their KK modes, i. e.,  
there exist only the zero modes and KK modes
of the $SU(3) \times SU(2) \times U(1)$ gauge fields and doublet Higgs
fields.
So, the extra dimensions can be large and the gauge hierarchy problem can
be solved in the mean time~\cite{AADD} because there does not exist the
proton 
decay from exchange
the $SU(5)$ gauge fields $X$ and $Y$ or the triplet Higgs at
all~\cite{ABK}.
Moreover, we can put the two Higgs doublets on the observable brane
instead of putting
two Higgs 5-plets in the bulk.  Furthermore, the wrong prediction of
the first and second generation mass ratio in the usual 4-dimensional
$SU(5)$
can be avoided, but, we do not have the charge quantization. 
In short, in this scenario, we not only
break the $SU(5)$ gauge symmetry, but also solve the triplet-doublet 
splitting problem, the gauge hierarchy problem, the proton decay problem,
and avoid
the wrong prediction of the first and second generation mass ratio in
the non-supersymmetric $SU(5)$ model, however, we lose the charge
quantization. 
By the way, the low energy unification and electroweak symmetry breaking 
 in this scenario deserve further study.

Furthermore, we generalize our models to the models with 7 or
higher dimensions where the extra space orbifolds are 
$T^n/(Z_2 \times Z_2')$ and $T^n/(Z_2)^n$, and comment on the models with
 bulk fermions and supersymmetry.
 
\section{SU(5) Models on $M^4 \times T^2/(Z_2 \times Z_2^{\prime})$}
In this section, we would like to discuss two models with
 SU(5) gauge symmetry breaking
on the space-time $M^4 \times T^2/(Z_2 \times Z_2^{\prime})$. We consider 
the 6-dimensional space-time which can be factorized into a product of the 
ordinary 4-dimensional space-time $M^4$ and the orbifold 
$T^2/(Z_2 \times Z_2^{\prime})$. The corresponding
coordinates are $x^{\mu}$, ($\mu = 0, 1, 2, 3$),
$y\equiv x^5$ and $z \equiv x^6$.
The radii for the circles along $y$ direction and $z$ direction are
$R_1$ and $R_2$, respectively.  Moreover, we assume that
the SU(5) gauge fields and two 5-plet Higgs fields ($H_u$ and
$H_d$) live on the whole space-time, and the Standard Model fermions
are on the observable brane, which is located at the
fixed point or line, although one can also discuss the models in which the
matters are
in the bulk or the Higgs fields are on the observable brane.

\subsection{Model I}
The first model is the generalization of previous
models~[$1-10$]. The orbifold $T^2/Z_2$ is obtained by $T^2$ moduloing 
the equivalent
class: $(y, z) \sim (-y, -z)$. We obtain the
orbifold $ T^2/(Z_2 \times Z_2^{\prime})$ by
 defining $y' \equiv y-\pi R_1/2$ 
and $z' \equiv z- \pi R_2/2$, and then, moduloing the equivalence class: 
$(y', z') \sim (-y', -z')$ further. The three non-equivalent
fixed points are $(y=0, z=0),$ $(y=0, z=\pi R_2)$, and 
$(y=\pi R_1/2, z=\pi R_2/2)$.

For a generic bulk field $\phi(x^{\mu}, y, z)$,
we can define two parity operators $P$ and $P'$ for the $Z_2$ and
$Z_2'$ symmetries, respectively
\begin{eqnarray}
\phi(x^{\mu},y, z)&\to \phi(x^{\mu},-y, -z )=P\phi(x^{\mu},y, z)
 ~,~\,
\end{eqnarray}
\begin{eqnarray}
\phi(x^{\mu},y', z')&\to \phi(x^{\mu},-y', -z' )=P' \phi(x^{\mu},y', z')
 ~.~\,
\end{eqnarray}

Denoting the field with ($P$, $P'$)=($\pm, \pm$) by 
$\phi_{\pm \pm}$, we obtain the following KK mode expansions
\begin{eqnarray}
  \phi_{++} (x^\mu, y, z) &=& {1\over\displaystyle
 {\pi \sqrt { R_1 R_2}}}
\sum_{n=0}^{\infty} \sum_{m=0}^{\infty}
\left( {1\over\displaystyle {\sqrt {2^{\delta_{n, 0}} 2^{\delta_{m, 0}}}}}
\phi_{++}^{(2n, 2m)}(x^{\mu}) \cos({{2 n y}\over R_1} + {{2mz}\over R_2})
 \right.\nonumber\\&&\left.
+ \phi^{(2n+1, 2m+1 )}_{++}(x^\mu) \cos({{(2 n+1) y}\over R_1} +
 {{(2m+1)z}\over R_2})\right)
 ~,~\,
\end{eqnarray}
\begin{eqnarray}
  \phi_{+-} (x^\mu, y, z) &=& {1\over\displaystyle
 {\pi \sqrt { R_1 R_2}}}
\sum_{n=0}^{\infty} \sum_{m=0}^{\infty}
\left( {1\over\displaystyle {\sqrt {2^{\delta_{n, 0}}}}}
\phi_{+-}^{(2n, 2m+1)}(x^{\mu}) \cos({{2 n y }\over R_1} + {{(2m+1)z}\over
R_2})
 \right.\nonumber\\&&\left.
+ \phi^{(2n+1, 2m )}_{+-}(x^\mu) {1\over\displaystyle {\sqrt
{2^{\delta_{m, 0}}}}}
 \cos({{(2 n+1) y}\over R_1} +
 {{2mz}\over R_2})\right)
 ~,~\,
\end{eqnarray}
\begin{eqnarray}
  \phi_{-+} (x^\mu, y, z) &=& {1\over\displaystyle
 {\pi \sqrt { R_1 R_2}}}
\sum_{n=0}^{\infty} \sum_{m=0}^{\infty}
\left( {1\over\displaystyle {\sqrt {2^{\delta_{n, 0}}}}}
\phi_{-+}^{(2n, 2m+1)}(x^{\mu}) \sin({{2 n y }\over R_1} + {{(2m+1)z}\over
R_2})
 \right.\nonumber\\&&\left.
+ \phi^{(2n+1, 2m )}_{-+}(x^\mu) {1\over\displaystyle {\sqrt
{2^{\delta_{m, 0}}}}}
 \sin({{(2 n+1) y}\over R_1} +
 {{2mz}\over R_2})\right)
 ~,~\,
\end{eqnarray}
\begin{eqnarray}
  \phi_{--} (x^\mu, y, z) &=& {1\over\displaystyle
 {\pi \sqrt { R_1 R_2}}}
\sum_{n=0}^{\infty} \sum_{m=0}^{\infty}
\left( {1\over\displaystyle {\sqrt {2^{\delta_{n, 0}} 2^{\delta_{m, 0}}}}}
\phi_{--}^{(2n, 2m)}(x^{\mu}) \sin({{2 n y}\over R_1} + {{2mz}\over R_2})
 \right.\nonumber\\&&\left.
+ \phi^{(2n+1, 2m+1 )}_{--}(x^\mu) \sin({{(2 n+1) y}\over R_1} +
 {{(2m+1)z}\over R_2})\right)
 ~,~\,
\end{eqnarray}
where $n, m$ are non-negative integers.
 Zero modes are contained only in $\phi_{++}$ fields,
so that the matter content of the massless sector is smaller
than that of the full 6-dimensional multiplet.
Moreover,  only $\phi_{++}$ and $\phi_{+-}$ fields have non-zero
values at $(y=0, z=0)$ and $(y=0, z=\pi R_2)$, and only 
 $\phi_{++}$ and $\phi_{-+}$ fields have non-zero
values at $(y=\pi R_1/2, z=\pi R_2/2)$.
 
We choose the following matrix representations of the parity assignments
which are expressed in the adjoint representaion of SU(5)
\begin{equation}
P={\rm diag}(+1, +1, +1, +1, +1)
~,~P'={\rm diag}(-1, -1, -1, +1, +1)
 ~.~\,
\end{equation}
So, upon the $P$ and $P'$ parities,
 the gauge generators $T^A$ where A=1, 2, ..., 24 for SU(5)
are separated into two sets: $T^a$ are the gauge generators for
the Standard Model gauge group, and $T^{\hat a}$ are the other broken
gauge generators 
\begin{equation}
P~T^a~P^{-1}= T^a ~,~ P~T^{\hat a}~P^{-1}= T^{\hat a}
~,~\,
\end{equation}
\begin{equation}
P'~T^a~P^{'-1}= T^a ~,~ P'~T^{\hat a}~P^{'-1}= - T^{\hat a}
~.~\,
\end{equation}

\renewcommand{\arraystretch}{1.4}
\begin{table}[t]
\caption{Parity assignment and masses ($n\ge 0, m \ge 0$) of the fields in
the SU(5) 
 gauge and Higgs multiplets for model I.
The index $a$ labels the unbroken $SU(3)\times SU(2) \times U(1)$ 
gauge generators, while $\hat a$
labels the other broken SU(5) gauge generators.
The indices $D$, $T$ are for doublet and triplet, respectively. 
\label{tab:chiral}}
\vspace{0.4cm}
\begin{center}
\begin{tabular}{|c|c|c|}
\hline        
$(P,P')$ & field & mass\\ 
\hline
$(+,+)$ &  $A^a_{\mu}$, $H^D_u$, $H^D_d$ & $\sqrt {(2n)^2/R_1^2+
(2m)^2/R_2^2}$ {\rm or}
$\sqrt {(2n+1)^2/R_1^2+ (2m+1)^2/R_2^2}$ \\
\hline
$(+,-)$ &  $A^{\hat{a}}_{\mu}$,  $H^T_u$, $H^T_d$ & $\sqrt
{(2n)^2/R_1^2+(2m+1)^2/R_2^2}$ 
{\rm or} $\sqrt {(2n+1)^2/R_1^2+(2m)^2/R_2^2}$ \\
\hline
$(-,+)$ &  $A^{\hat{a}}_5$, $A^{\hat{a}}_6$ 
& $\sqrt {(2n)^2/R_1^2+(2m+1)^2/R_2^2}$ 
{\rm or} $\sqrt {(2n+1)^2/R_1^2+(2m)^2/R_2^2}$ \\
\hline
& & $\sqrt {(2n)^2/R_1^2+ (2m+2)^2/R_2^2}$ {\rm or}\\
$(-,-)$ &  $A^{a}_5$, $A^{a}_6$ &  $\sqrt {(2n+2)^2/R_1^2+ (2m)^2/R_2^2}$
{\rm or}\\
& & $\sqrt {(2n+1)^2/R_1^2+(2m+1)^2/R_2^2}$ \\
\hline
\end{tabular}
\end{center}
\end{table}

\renewcommand{\arraystretch}{1.4}
\begin{table}[t]
\caption{For the model I, the gauge fields, Higgs fields and gauge group
on
 the 3-branes which are located
at the fixed points $(y=0, z=0),$ $(y=0, z=\pi R_2)$, and 
$(y=\pi R_1/2, z=\pi R_2/2)$.
\label{tab:chiral}}
\vspace{0.4cm}
\begin{center}
\begin{tabular}{|c|c|c|}
\hline        
Brane position & field & gauge group\\ 
\hline
$(y=0, z=0)$ & $A_{\mu}^A$, $H_u$, $H_d$ & $SU(5)$ \\
\hline
$(y=0, z=\pi R_2)$ & $A_{\mu}^A$, $H_u$, $H_d$ & $SU(5)$ \\
\hline
$(y=\pi R_1/2, z=\pi R_2/2)$ & $A_{\mu}^a$, $H_u^D$, $H_d^D$,
$A^{\hat{a}}_5$, $A^{\hat{a}}_6$ & $SU(3)\times SU(2)\times U(1) $ \\
\hline
\end{tabular}
\end{center}
\end{table}

It is not difficult to obtain the particle spectra in the model, which are
given in 
Table 1. And the gauge fields, Higgs fields, and gauge group on the
3-brane which is
located at the fixed point are given in Table 2. If the Standard Model
fermions were on the
3-brane which is located at $(0, 0)$ or $(0, \pi R_2)$, the gauge group
will be $SU(5)$,
then, the SM fermions must form the full $10^i + \bar 5^i$ $SU(5)$
multiplets. In order
to avoid the proton decay, the GUT scale ($1/R_1$ or $1/R_2$) must be the
order of
 $10^{15\sim 16}$ GeV, and we have to put the two Higgs 5-plets in the
bulk to
give the GUT scale masses to the Higgs triplets. So, there may exist the
gauge hierarchy problem.

If the Standard Model fermions were on the brane at $(\pi R_1/2, \pi
R_2/2)$,
the gauge group will be $SU(3)\times SU(2)\times U(1)$. If we considered
the
SM fermions form the full $10^i + \bar 5^i$ $SU(5)$ multiplets, the proton
may decay
by exchange $A_5^{\hat a}$ or $A_6^{\hat a}$, so, the GUT scale must be at
order of $10^{15\sim 16}$ GeV and we have to put the two Higgs 5-plets in
the bulk.
Thus, we might have the gauge hierarchy problem. If we considered the
SM fermions do not form the full $SU(5)$ multiplets and they only preserve
the
$SU(3)\times SU(2)\times U(1)$ gauge symmetry, the proton decay problem
will be
avoided because the SM fermions do not couple to the gauge fields 
related to the broken gauge generators. 
In addition, we can put the two Higgs doublets on the observable brane
instead of putting
two Higgs 5-plets in the bulk. Therefore, the extra dimensions might be
large 
and the gauge hierarchy problem might be solved in the mean
time~\cite{AADD}. 

\subsection{Model II}
The second model is different from above
model. The orbifold $T^2/(Z_2\times Z_2^{\prime})$ is
 obtained by $T^2$ moduloing the equivalent
classes: $y \sim -y$ and $ z \sim -z$.  The four fixed points are
$(y=0, z=0),$ $(y=0, z=\pi R_2/2),$ $(y=\pi R_1/2, z=0)$, and 
$(y=\pi R_1/2, z=\pi R_2/2)$. And the four fixed lines are
$y=0$, $y=\pi R_1/2$, $z=0$, $z=\pi R_2/2$.

For a generic bulk field $\phi(x^{\mu}, y, z)$,
we can define two parity operators $P$ and $P'$ for the $Z_2$ and
$Z_2'$ symmetries, respectively
\begin{eqnarray}
\phi(x^{\mu},y, z)&\to \phi(x^{\mu},-y, z )=P\phi(x^{\mu},y, z)
 ~,~\,
\end{eqnarray}
\begin{eqnarray}
\phi(x^{\mu},y, z)&\to \phi(x^{\mu},y, -z )=P' \phi(x^{\mu},y, z)
 ~.~\,
\end{eqnarray}

Denoting the field with ($P$, $P'$)=($\pm, \pm$) by 
$\phi_{\pm \pm}$, we obtain the following KK mode expansions
\begin{eqnarray}
  \phi_{++} (x^\mu, y, z) &=&
{1\over\displaystyle
 {\pi \sqrt {R_1 R_2}}}
\sum_{n=0}^{\infty} \sum_{m=0}^{\infty}
 {1\over\displaystyle
 { \sqrt {2^{\delta_{n, 0}} 2^{\delta_{m, 0}}}}}
 \phi^{(n, m )}_{++}(x^\mu) 
\nonumber\\&&
\cos{ {n y} \over {R_1}} \cos{ {m z} \over {R_2}}
 ~,~\,
\end{eqnarray}
\begin{eqnarray}
  \phi_{+-} (x^\mu, y, z) &=& {1\over\displaystyle
 {\pi \sqrt {R_1 R_2}}}
\sum_{n=0}^{\infty} \sum_{m=1}^{\infty}
{1\over\displaystyle
 { \sqrt {2^{\delta_{n, 0}}}}}
 \phi^{(n, m )}_{+-}(x^\mu) 
\cos{{n y} \over {R_1}} 
\nonumber\\&&
\sin{{m z} \over {R_2}}
 ~,~\,
\end{eqnarray}
\begin{eqnarray}
  \phi_{-+} (x^\mu, y, z) &=& 
{1\over\displaystyle
 {\pi \sqrt {R_1 R_2}}}
\sum_{n=1}^{\infty} \sum_{m=0}^{\infty}
{1\over\displaystyle
 {\sqrt {2^{\delta_{m, 0}}}}}
 \phi^{(n, m )}_{-+}(x^\mu) 
\sin{{n y} \over {R_1}} 
\nonumber\\&&
\cos{ {mz} \over {R_2}}
 ~,~\,
\end{eqnarray}
\begin{eqnarray}
  \phi_{--} (x^\mu, y, z) &=& {1\over\displaystyle
 {\pi \sqrt {R_1 R_2}}}
\sum_{n=1}^{\infty} \sum_{m=1}^{\infty}
 \phi^{(n, m)}_{--}(x^\mu) 
\sin{ {n y} \over {R_1}} 
\nonumber\\&&
\sin{{m z} \over {R_2}}
 ~,~\,
\end{eqnarray}
where $n, m$ are non-negative integer, and the 4-dimensional
fields $\phi^{(n, m )}_{++}$, $\phi^{(n, m )}_{+-}$,
$\phi^{(n, m )}_{-+}$ and $\phi^{(n, m )}_{--}$
have masses $\sqrt {n^2/R_1^2+m^2/R_2^2}$ upon the
compactification. Because only $\phi_{++}$ fields have zero modes,
 the matter content of the massless sector is smaller
than that of the full 6-dimensional multiplet, too.
In addition,  only $\phi_{++}$ fields have non-zero
values on the 3-branes at the fixed points, only
$\phi_{++}$ and $\phi_{+-}$ have non-zero values on the
4-branes at the fixed lines $y=0$ and $y=\pi R_1/2$,
and only $\phi_{++}$ and $\phi_{-+}$ have non-zero values on the
4-branes at the fixed lines $z=0$ and $z=\pi R_2/2$.

We choose the following matrix representations of the parity assignments
which are expressed in the adjoint representaion of SU(5)
\begin{equation}
P={\rm diag}(-1, -1, -1, +1, +1)
~,~P'={\rm diag}(-1, -1, -1, +1, +1)
 ~.~\,
\end{equation}
Then, upon the $P$ and $P'$ parities,
the gauge generators $T^A$ where A=1, 2, ..., 24 for SU(5)
are also separated into two sets: $T^a$ are the gauge generators for
the Standard Model gauge group, and $T^{\hat a}$ are the other broken 
gauge generators
\begin{equation}
P~T^a~P^{-1}= T^a ~,~ P~T^{\hat a}~P^{-1}= - T^{\hat a}
~,~\,
\end{equation}
\begin{equation}
P'~T^a~P^{'-1}= T^a ~,~ P'~T^{\hat a}~P^{'-1}= - T^{\hat a}
~.~\,
\end{equation}

\renewcommand{\arraystretch}{1.4}
\begin{table}[tt]
\caption{Parity assignment and masses ($n \ge 0, m \ge 0$ ) of the fields
in the SU(5) gauge
 and Higgs multiplets for model II.
\label{tab:chiralI}}
\vspace{0.4cm}
\begin{center}
\begin{tabular}{|c|c|c|}
\hline    
$(P,P')$ & field & mass\\ 
\hline
$(+,+)$ &  $A^a_{\mu}$, $H^D_u$, $H^D_d$ & $\sqrt {n^2/R_1^2+m^2/R_2^2}$\\
\hline
$(+,-)$ &  $A^{\hat{a}}_{5}$, $A^{a}_6$ & $\sqrt
{n^2/R_1^2+(m+1)^2/R_2^2}$ \\
\hline
$(-,+)$ &  $A^{a}_5$, $A^{\hat{a}}_6$ 
& $\sqrt {(n+1)^2/R_1^2+m^2/R_2^2} $\\
\hline
$(-,-)$ &  $A^{{\hat a}}_{\mu}$, $H^T_u$, $H^T_d$ & 
$\sqrt {(n+1)^2/R_1^2+(m+1)^2/R_2^2} $ \\
\hline
\end{tabular}
\end{center}
\end{table}

\renewcommand{\arraystretch}{1.4}
\begin{table}[t]
\caption{For the model II, the gauge fields, Higgs fields and gauge group
on
 the 3-branes which are located
at the fixed points $(y=0, z=0),$ $(y=0, z=\pi R_2/2)$, 
$(y=0, z=\pi R_2/2)$ and $(y=\pi R_1/2, z=\pi R_2/2)$, and
on the 4-branes which are located at $y=0$, $z=0$, $y=\pi R_1/2$
and $z=\pi R_2/2$.
\label{tab:chiral}}
\vspace{0.4cm}
\begin{center}
\begin{tabular}{|c|c|c|}
\hline        
Brane position & field & gauge group\\ 
\hline
$(y=0, z=0)$ & $A_{\mu}^a$, $H_u^D$, $H_d^D$ & $SU(3)\times SU(2)\times
U(1)$ \\
\hline
$(y=0, z=\pi R_2/2)$ & $A_{\mu}^a$, $H_u^D$, $H_d^D$ & $SU(3)\times
SU(2)\times U(1)$ \\
\hline
$(y=\pi R_1/2, z=0)$ & $A_{\mu}^a$, $H_u^D$, $H_d^D$ & $SU(3)\times
SU(2)\times U(1)$ \\
\hline
$(y=\pi R_1/2, z=\pi R_2/2)$ & $A_{\mu}^a$, $H_u^D$, $H_d^D$  & 
$SU(3)\times SU(2)\times U(1) $ \\
\hline
$y=0$ or $y=\pi R_1/2$ &  $A_{\mu}^a$, $H_u^D$, $H_d^D$,
$A^{\hat{a}}_{5}$, $A^{a}_6$
& $SU(3)\times SU(2)\times U(1)$ \\
\hline
$z=0$  or $z=\pi R_2/2$ & $A_{\mu}^a$, $H_u^D$, $H_d^D$, $A^{a}_{5}$,
$A^{\hat{a}}_6$
& $SU(3)\times SU(2)\times U(1)$ \\
\hline
\end{tabular}
\end{center}
\end{table}

The particle spectra in this model are given in 
Table 3, and the gauge fields, Higgs fields, and gauge group on the
3-brane which is
located at the fixed point or on the 4-brane which is located at the
fixed line are given in Table 2.

If the Standard Model fermions were on the 4-brane which is located at
$y=0$, or $z=0$, or $y=\pi R_1/2$, or $z=\pi R_2/2$,
the gauge symmetry will be $SU(3)\times SU(2)\times U(1)$. And the 
phenomenological discussions will be similar to those in Model I when we
put
the SM fermions on the 3-brane at $(\pi R_1/2,\pi R_2/2 )$, except that,
if
we considered the SM fermions form the full $10^i + \bar 5^i$ $SU(5)$
multiplets,
we will double the generations and each generation in the SM comes from
the zero modes of
two generations on the 4-brane due to the projection operator $P^y$ or
$P^z$.

If the Standard Model fermions were on the 3-brane which is located
at any fixed point, the gauge symmetry will be
$SU(3)\times SU(2)\times U(1)$. In addition, on the observable 3-brane,
 we project out not only the zero 
modes of the gauge fields related to the broken SU(5) gauge generators and
the
triplet Higgs fields, but also their KK modes, i. e.,  
there exist only the zero modes and KK modes
of the $SU(3) \times SU(2) \times U(1)$ gauge fields and doublet Higgs
fields.
So, the extra dimensions can be large and the gauge hierarchy problem can
be solved in the mean time~\cite{AADD} because there does not exist the
proton 
decay from exchange
the $SU(5)$ gauge fields $X$ and $Y$ or the triplet Higgs at
all~\footnote{One
might worry about the proton decay mediated by the virtual black hole and
 wormhole background from quantum gravity~\cite{KANE}. However, with our
present
knowledge of black hole and wormhole physics, the violation of
 global symmetry by gravity is extremely suppressed, even if the
fundamental
Planck scale was as low as a few TeV~\cite{ABK}.}. 
Moreover, we can put the two Higgs doublets on the observable brane
instead of putting
two Higgs 5-plets in the bulk.  Furthermore, the wrong prediction of
the first and second generation mass ratio in the usual 4-dimensional
$SU(5)$
can be avoided, but, we do not have the charge quantization. 
In short, in this scenario, we not only
break the $SU(5)$ gauge symmetry, but also solve the triplet-doublet 
splitting problem, the gauge hierarchy problem, the proton decay problem,
and avoid
the wrong prediction of the first and second generation mass ratio in
the non-supersymmetric $SU(5)$ model, however, we lose the charge
quantization.
 By the way, the low energy unification and electroweak symmetry breaking 
 in this scenario are under investigation.

\section{Generalization Discussion and Conclusion} 
It is easy to generalize our models to the models with $4+n$ dimensions
where $n > 2$ and the original
extra space manifold is a n-torii, i. e., $T^n$. 
We assume that the coordinates for the extra dimensions
are $(y^1, y^2, ..., y^n)$, the SU(5) gauge fields and two 5-plet Higgs 
fields are in the bulk, and the observable brane is located at the fixed
point or
hypersurface. We obtain the first kind of model on the space-time
$M^4\times T^n/({Z_2\times Z_2'})$ by $T^n$ moduloing the equivalent
classes:
$(y^1, y^2, ..., y^n) \sim (-y^1, -y^2, ..., -y^n)$,
$(y^{'1}$, $y^{'2}$, ..., $y^{'n}$) $\sim$ ($-y^{'1}$, 
$-y^{'2}$, ..., $-y^{'n}$),
whose corresponding parity operators are $P$ and $P'$,
respectively. Under SU(5) gauge symmetry, we choose the
representations for $P$ and $P'$ as 
$P$={\rm diag} (+1, +1, +1, +1, +1), and
$P'$={\rm diag} ($-1, -1, -1,$ +1, +1), so,
$P$$T^a$$P^{-1}$= $T^a$, $P$$T^{\hat a}$$P^{-1}$=$T^{\hat a}$,
$P'$$T^a$$P^{'-1}$= $T^a$, $P$$T^{\hat a}$$P^{'-1}$=$-T^{\hat a}$.
Then, for the zero modes, the SU(5) gauge symmetry is broken 
down to the Standard Model gauge symmetry.
Furthermore, we can obtain the second kind of model on the space-time
$T^n/(Z_2)^n$ by $T^n$ moduloing the equivalent classes:
$y^1 \sim -y^1$, $y^2 \sim -y^2$, ..., and $y^n \sim -y^n$, whose
corresponding parity operators are $P^1$, $P^2$, ...,
$P^n$. The SU(5) representations of $P^1$, $P^2$, ...,
$P^n$ are
$P^i={\rm diag} (-1, -1, -1, +1, +1)$ for $i=1, 2, ..., n$.
So, $P^i T^a P^{-i}= T^a, P^i T^{\hat a} P^{-i} = - T^{\hat a}$.
Then, for the zero modes, the SU(5) gauge symmetry is broken 
down to the $SU(3)\times SU(2)\times U(1)$ gauge symmetry.
In particular, on the observable 3-brane which is located at any one 
of the $2^n$ fixed points, there exist only the zero modes
and KK modes of the $SU(3) \times SU(2) \times U(1)$ gauge fields because
 the zero modes and KK modes of the gauge fields related to
the broken $SU(5)$ gauge generators are projected out.
The phenomenology discussions are similar to those in section 2.

Furthermore, we can consider the models with bulk fermions, and 
we have to add extra $SU(5)$ fermion multipets due to the 
anomaly cancellation~\cite{BGT}. For instance, the models, which contain
three left-handed singlets, three $5$, six $\bar 5$, six $10$, 
five ${\bar {10}}$ fermion multiplets under $SU(5)$ gauge symmetry, and
a self-dual antisymmetric tensor, is anomaly free~\cite{BGT}. If we
considered
the extra space orbifold $T^2/(Z_2)^3$ or
$T^2/(Z_2)^4$, the Standard Model fermions
come from the zero modes of six $\bar 5$ and six $10$ multiplets,
and we project out the zero modes of all the other fermion
multiplets~\cite{LY}.
The reason why we consider the extra space orbifold $T^2/(Z_2)^3$ or
$T^2/(Z_2)^4$ is that
the 6-dimensional fermions contain 4-dimensional fermions and their
mirrors,
and we want to project out the zero modes of those mirror particles.
And the reason why we need six $\bar 5$ and six $10$ $SU(5)$ multiplets is
that
we have $Z_2$ projections on the zero modes of all the multiplets. Of
course,
the gauge hierarchy problem, gauge coupling unification and 
radiative electroweak symmetry breaking in those non-supersymmetric GUT
models deserve further investigation~\cite{LY}.

Another natural generalization is supersymmetry. If we considered the
$N=1$
6-dimensional supersymmetry, the gauginos 
(and gravitinos) have positive chirality and the bulk hypermultiplets have
negative
chirality. Therefore, we can put the Standard Model fermions on
the brane, and add suitable bulk hypermultiplets to cancel the
anomaly from the gauge multiplet, for example, we can add a $24$ multiplet
in the adjoint representation of $SU(5)$ or ten $5+\bar 5$ $SU(5)$
multiplets, and
the Higgs fields will be on the brane or from the ten 
$5+\bar 5$ multiplets, repectively~\cite{TWTY}. If we considered the $N=2$ 
6-dimensional supersymmetry,
the theory will be anomaly free, and we have to put the Standard Model
fermions on the brane because we can not have the hypermultiplets in the
bulk~\cite{SUSY}.

In short, we construct two SU(5) GUT models on the space-time
$M^4 \times T^2/(Z_2 \times Z_2^{\prime})$  where the SU(5)
 gauge and Higgs fields 
are in the bulk and the Standard Model fermions are on the observable
brane at the fixed point or line.
For the zero modes, the SU(5) gauge symmetry is broken down
to the $SU(3) \times SU(2) \times U(1) $ gauge symmetry due to
 non-trivil orbifold projection.
In particular, if we put the Standard Model fermions on the 3-brane at the
fixed point
in Model II, we only have the zero modes and KK modes of the
Standard Model gauge fields and two Higgs doublets on the observable
3-brane.
 So, we can not only break the $SU(5)$ gauge symmetry, but also solve the
triplet-doublet 
splitting problem, the gauge hierarchy probelm, the proton decay problem,
and avoid
the wrong prediction of the first and second generation mass ratio in
the non-supersymmetric $SU(5)$ model, although we lose the charge
quantization. 

\section*{Acknowledgments}
This research was supported in part by the U.S.~Department of Energy under
 Grant No.~DOE-EY-76-02-3071.

\end{document}